\documentclass[letterpaper, 10 pt, conference]{ieeeconf}
\IEEEoverridecommandlockouts    

\usepackage{multirow}
\usepackage{lipsum}
\usepackage{amsmath}
\usepackage{amssymb}
\usepackage[T1]{fontenc}
\usepackage[ruled,vlined]{algorithm2e}
\usepackage{graphicx}
\graphicspath{{./Figures/}}
\usepackage{float}
\usepackage{booktabs}
\usepackage{balance}

\title{\LARGE \bf Impact-Driven Quantum Decomposition for Traffic Zone Partitioning: A Hybrid Gate-Model Framework}

\author{
 	\parbox{\textwidth}{%
 		\centering
 		Ruimin Ke$^{1*}$, Talha Azfar$^{1}$, Kaicong Huang$^{1}$, Shuyang Li$^{1}$
 	}%
 	\thanks{$^{1}$Department of Civil and Environmental Engineering, Rensselaer Polytechnic Institute, Troy, NY, USA
 		{\tt\small ker@rpi.edu, azfart@rpi.edu, huangk10@rpi.edu, lis36@rpi.edu}}%
 }

\begin{document}
	
	\maketitle
    \vspace{-6pt}
	\thispagestyle{empty}
	\pagestyle{empty}

\begin{abstract}
Partitioning transportation networks into balanced and spatially coherent traffic zones is a fundamental yet computationally challenging task in intelligent transportation systems. The resulting optimization problem exhibits dense interactions among decision variables and can be formulated as a Quadratic Unconstrained Binary Optimization (QUBO) model. While quantum optimization naturally aligns with such quadratic energy representations, current noisy intermediate-scale quantum hardware imposes limitations on problem size, connectivity, and circuit reliability. This paper proposes an impact-driven hybrid quantum--classical optimization framework for traffic zone partitioning that bridges transportation-scale optimization models and practical gate-based quantum processors. Instead of static geographic decomposition, the method estimates the energy impact of decision variables and selectively assigns quantum computation to influential subproblems while a classical coordination loop maintains global feasibility. The framework is implemented using the Iskay optimizer and evaluated on the IBM Quantum System One backend. Experiments compare direct quantum optimization, classical iterative SubQUBO refinement, and the proposed hybrid approach. Results show that impact-guided decomposition improves convergence behavior and produces more coherent spatial partitions relative to classical refinement, while remaining consistent with hardware constraints. Although the hybrid method does not outperform the best direct quantum solution, it demonstrates a practical pathway toward scalable hybrid optimization for transportation applications under current quantum hardware conditions.
\end{abstract}

\section{Introduction}

Modern intelligent transportation systems increasingly rely on spatial partitioning of urban networks into operational traffic zones~\cite{li2019research,yang2007method}. Such zoning supports a wide range of applications including demand estimation, service allocation, signal coordination, and regional performance monitoring. Effective partitioning must simultaneously satisfy multiple requirements: balanced workload distribution, geographic coherence, and operational interpretability.

As transportation networks grow in scale and heterogeneity, these requirements produce combinatorial decision spaces whose complexity increases exponentially with system size. Classical exact optimization methods, including mixed-integer programming formulations, become computationally intractable for large metropolitan networks~\cite{zhou2025flow}. Heuristic and clustering approaches offer practical alternatives but often struggle to reconcile globally coupled balance constraints with local spatial structure~\cite{dixit2023quantum}. Well-established graph partitioning and multilevel refinement methods remain highly effective for many practical zoning problems, particularly at moderate scales~\cite{karypis1998fast}.

Quadratic Unconstrained Binary Optimization (QUBO)~\cite{lewis2017quadratic} provides a unified representation capable of encoding both global and spatial objectives within a single energy function~\cite{glover2018tutorial}. This representation naturally aligns with emerging quantum optimization paradigms, which operate by exploring correlated configurations of interacting variables. However, a fundamental mismatch remains between transportation-scale optimization problems and the capabilities of current noisy intermediate-scale quantum (NISQ) hardware~\cite{azfar2025quantum}. Limited qubit connectivity, gate noise, and circuit depth constraints prevent direct encoding of realistic transportation systems~\cite{zhuang2024quantum}. Consequently, even when a transportation problem admits a QUBO formulation, larger transportation-scale instances may exceed the executable capacity of current quantum devices~\cite{azfar2026shallow,azfar2026hardware}.

\begin{figure*}
		\centering
		\includegraphics[width=\textwidth]{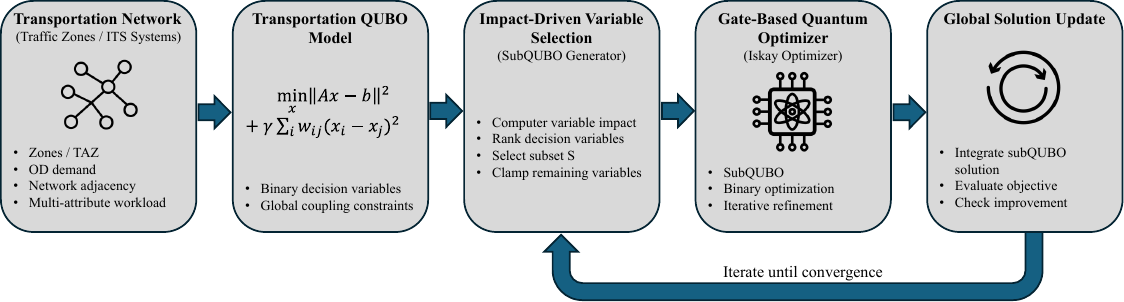}
		\caption{Impact-driven hybrid quantum–classical optimization framework for traffic zone partitioning. Transportation network data are formulated as a QUBO model, influential variables are selected through energy sensitivity analysis, reduced subQUBO problems are optimized using a gate-based quantum solver, and solutions are iteratively integrated into a global zoning configuration until convergence.}
		\label{fig:framework}
	\end{figure*}

To address this scale mismatch, this work proposes an impact-driven hybrid quantum--classical optimization framework for traffic zone partitioning. Instead of decomposing the transportation network using static geographic partitions, the proposed approach dynamically identifies influential decision variables based on energy sensitivity and selectively applies quantum optimization only to reduced subQUBO instances while maintaining global coordination through classical updates. By restricting quantum execution to influential variable subsets, the framework enables optimization of transportation-scale problems whose full QUBO representations exceed available quantum hardware resources. The workflow of the proposed framework is illustrated in Fig.~\ref{fig:framework}.

The contributions of this paper are threefold:

1) A transportation-aware QUBO formulation that captures workload balance and spatial coherence within a unified energy framework;

2) An impact-driven decomposition strategy derived from energy sensitivity analysis that targets a scalable pathway to traffic zone partitioning through selective quantum subproblem execution;

3) A hybrid gate-model implementation using the Iskay optimizer evaluated on the IBM Quantum System One backend, providing empirical insight into hybrid optimization behavior under realistic hardware constraints.

This work does not aim to demonstrate quantum computational advantage; instead, it investigates how hybrid algorithm design can reconcile transportation-scale optimization structure with practical hardware constraints in the current NISQ era. 
This paper positions traffic zone partitioning as a representative transportation application through which hybrid quantum optimization design principles can be studied under realistic system constraints.
As an initial conference investigation, the primary objective is to establish the formulation and hybrid computational workflow under practical hardware constraints rather than to claim operational superiority over mature classical partitioning algorithms.

The remainder of the paper develops the theoretical foundations of the model, introduces the impact-driven algorithm, and evaluates its behavior through controlled experimental validation.

\section{Theoretical Foundations and Design Logic}

The development of scalable optimization methods for transportation systems requires alignment between physical system structure and computational representation. Traffic zone partitioning provides a setting in which operational planning objectives, spatial constraints, and computational complexity interact. The framework proposed in this study is derived through a sequence of abstractions that transform a transportation planning problem into a form compatible with hybrid quantum optimization.

Consider a transportation network represented by a spatial graph
\begin{equation}
G = (V,E),
\label{eq:graph}
\end{equation}
where each node corresponds to a traffic zone and edges represent spatial adjacency or interaction relationships. A zoning configuration assigns each zone to an operational region, producing a system level state whose quality depends simultaneously on workload balance and geographic coherence.

Let
\begin{equation}
x = (x_1, x_2, \dots, x_n)^{T}
\label{eq:state_vector}
\end{equation}
denote the binary decision vector describing zone assignments. Each configuration represents one possible operational partition of the transportation system.

\subsection{Energy Based Interpretation of Transportation Partitioning}

Transportation zoning objectives combine multiple competing requirements. Balanced workloads require redistribution across the entire network, whereas spatial coherence encourages neighboring zones to remain grouped.

To unify these objectives, the zoning problem is interpreted through an energy representation
\begin{equation}
H(x) = H_{\text{balance}}(x) + H_{\text{spatial}}(x),
\label{eq:energy_decomposition}
\end{equation}
where lower energy corresponds to improved operational quality.

Equation~(\ref{eq:energy_decomposition}) transforms transportation planning into an energy minimization problem defined over discrete configurations. The resulting landscape contains numerous locally optimal states separated by interaction barriers arising from conflicting objectives.

\subsection{Interaction Structure and Computational Difficulty}

Unlike classical graph partitioning problems that depend primarily on topology~\cite{kernighan1970efficient}, transportation systems introduce attribute driven coupling. The contribution of a single zone to regional balance depends on aggregated system totals, implying that distant zones may influence each other indirectly.

This interaction structure can be expressed generically as
\begin{equation}
H(x) = \sum_i h_i x_i + \sum_{i<j} J_{ij} x_i x_j + C,
\label{eq:quadratic_energy}
\end{equation}
where coefficients $h_i$ represent individual zone effects and $J_{ij}$ capture pairwise interactions induced by balance and spatial constraints.

Equation~(\ref{eq:quadratic_energy}) reveals that transportation partitioning forms a dense quadratic optimization problem. Improvements often require coordinated modification of multiple variables, which explains the difficulty encountered by local search heuristics that adjust decisions sequentially.

\subsection{Implications for Optimization Strategy}

The quadratic interaction form in (\ref{eq:quadratic_energy}) implies that optimization difficulty is not uniformly distributed across variables. At any stage of optimization, certain zones contribute disproportionately to potential energy reduction. These zones correspond to regions where competing objectives remain unresolved.

Consequently, scalability depends on identifying influential components of the decision space rather than attempting uniform optimization across all variables simultaneously. This observation motivates a selective computational strategy in which optimization effort is dynamically allocated according to variable influence.

\subsection{Compatibility with Quantum Optimization}

Quantum optimization algorithms operate naturally on quadratic energy models such as (\ref{eq:quadratic_energy}). Through coherent evolution, quantum processes explore correlated configurations that update interacting variables jointly rather than sequentially.

However, present quantum processors cannot encode full scale transportation systems directly. The theoretical implication is therefore not that quantum computation replaces classical optimization, but that it should be applied selectively to portions of the energy landscape where correlated exploration is most beneficial.

This reasoning leads directly to an impact driven decomposition framework, in which the interaction structure identified in (\ref{eq:quadratic_energy}) determines how quantum resources are incorporated within a hybrid optimization workflow. The following section formalizes this idea through an adaptive algorithm that allocates quantum computation according to measured variable influence.

The following sections operationalize these theoretical observations into an executable optimization framework.

\section{Traffic Zone Partitioning as a QUBO Model}

The theoretical discussion established that transportation partitioning can be interpreted as an energy minimization problem defined over interacting spatial units~\cite{nishimori2001statistical}. This section formalizes that interpretation by deriving a quadratic unconstrained binary optimization representation for traffic zone partitioning. The formulation preserves transportation meaning while producing a structure compatible with hybrid quantum optimization.

\subsection{Binary Representation of Zone Assignment}

Consider a transportation system consisting of $n$ traffic zones divided into two operational regions. Each zone is associated with a binary decision variable

\begin{equation}
x_i =
\begin{cases}
1, & \text{zone } i \text{ assigned to region A},\\
0, & \text{zone } i \text{ assigned to region B}.
\end{cases}
\end{equation}

The vector $x=(x_1,\dots,x_n)$ defines a global zoning configuration.

\subsection{Balance Objective}

Each traffic zone possesses multiple operational attributes such as traffic volume, service workload, and activity intensity. Let

\begin{equation}
A \in \mathbb{R}^{n\times m}
\end{equation}

denote the attribute matrix, where $A_{ik}$ represents attribute $k$ of zone $i$.

Balanced partitioning requires regional totals to approach half of the system aggregate:

\begin{equation}
T_k=\frac{1}{2}\sum_{i=1}^{n}A_{ik}.
\end{equation}

Deviation from balance is measured through a quadratic penalty

\begin{equation}
H_B=
\sum_{k=1}^{m}
\left(
\sum_{i=1}^{n}A_{ik}x_i-T_k
\right)^2 .
\end{equation}

\subsection{Quadratic Expansion}

Expanding the balance term yields interactions among decision variables. To maintain column compatibility, the expansion is written in aligned form:

\begin{align}
H_B
= \sum_{k=1}^{m}
\Bigg(
&\sum_{i} A_{ik}^2 x_i
+ 2\sum_{i<j} A_{ik}A_{jk}x_ix_j
\nonumber\\
&-2T_k\sum_{i}A_{ik}x_i
\Bigg)
+ C .
\end{align}

This expression reveals that balance requirements introduce quadratic coupling between zones that may be geographically distant. Consequently, optimization cannot rely solely on local adjustments.

\subsection{Spatial Coherence Term}

Transportation regions are expected to remain spatially contiguous. Let adjacency weights $W_{ij}$ quantify interaction strength between neighboring zones. Spatial coherence is encouraged using

\begin{equation}
H_A=\lambda\sum_{i,j}W_{ij}(x_i-x_j)^2 .
\end{equation}

Using binary identities, this becomes

\begin{equation}
H_A=\lambda\sum_{i,j}W_{ij}
\left(x_i+x_j-2x_ix_j\right).
\end{equation}

This term promotes consistent assignments among adjacent zones while allowing balance objectives to influence global structure.

\subsection{Unified QUBO Formulation}

The total objective combines balance and spatial terms:

\begin{equation}
H(x)=H_B+H_A .
\end{equation}

Collecting coefficients produces the quadratic form

\begin{equation}
H(x)=x^{T}Qx + C,
\end{equation}

where matrix $Q$ encodes both attribute interactions and adjacency relationships.

Diagonal elements represent individual zone contributions, while off diagonal elements capture interaction tensions arising from balance and spatial coherence.

\subsection{Interpretation of Model Components}

\begin{table}
\caption{Interpretation of QUBO components}
\centering
\small
\setlength{\tabcolsep}{3.5pt}
\begin{tabular}{lll}
\toprule
Component & Mathematical role & Transportation meaning\\
\midrule
Linear term & node bias & workload contribution\\
Quadratic term & interaction coupling & spatial or balance influence\\
Constant term & offset & evaluation baseline\\
\bottomrule
\end{tabular}
\end{table}

The QUBO representation unifies heterogeneous transportation objectives into a single energy function while preserving interpretability of planning variables.

\subsection{Implication for Scalable Optimization}

An important consequence of the formulation is the density of quadratic interactions. Because balance constraints connect many zones simultaneously, solving the full optimization problem requires coordination across widely separated variables. This observation explains why static geographic decomposition is insufficient and motivates the adaptive optimization strategy introduced in the following section.
\section{Impact Driven Quantum Decomposition}

The QUBO formulation derived in the previous section reveals that traffic zone partitioning is governed by dense quadratic interactions among decision variables. Although this representation unifies transportation objectives into a single optimization framework, its dimensionality quickly exceeds the computational capacity of both classical exhaustive methods and current quantum hardware. The purpose of this section is therefore to develop a scalable solution strategy that preserves global interaction structure while restricting quantum computation to tractable subproblems.

Rather than decomposing the transportation network according to geographic boundaries, the proposed approach derives decomposition directly from the energy landscape defined by the QUBO model. The central idea is that only a subset of variables dominates the evolution of the objective function at any stage of optimization. Identifying these influential variables allows computational effort to be concentrated where improvement potential is greatest.

\subsection{Energy Sensitivity of Decision Variables}

Let the global objective be expressed as

\begin{equation}
H(x)=x^{T}Qx ,
\end{equation}

where $x$ is the binary zoning vector and $Q$ encodes balance and spatial interactions.

Consider flipping a single variable $x_i$. The resulting change in objective value is

\begin{equation}
\Delta H_i = H(x^{(i)})-H(x),
\end{equation}

where $x^{(i)}$ denotes the configuration obtained after inverting $x_i$.

Expanding the quadratic form gives

\begin{align}
\Delta H_i
= (1-2x_i)
\left(
Q_{ii}
+
2\sum_{j\neq i} Q_{ij}x_j
\right).
\end{align}

This quantity measures how strongly zone $i$ contributes to the current imbalance of the system. Large magnitudes indicate locations where coordinated reassignment may significantly reduce the objective value. We refer to this quantity as the variable impact.

\subsection{Adaptive Selection of Active Variables}

Let $q$ denote the number of decision variables that can be optimized within available quantum resources. At each iteration, variables are ranked according to their impact values, and the $q$ most influential variables form an active set $S$. The remaining variables constitute a fixed environment set $F$.

Reordering the QUBO matrix according to this partition yields

\begin{equation}
Q=
\begin{bmatrix}
Q_{SS} & Q_{SF}\\
Q_{FS} & Q_{FF}
\end{bmatrix}.
\end{equation}

Variables in $F$ remain fixed at their current assignments. Substituting these values into the objective produces a reduced optimization problem defined only over $x_S$:

\begin{equation}
H_S(x_S)
=
x_S^{T}Q_{SS}x_S
+
b_S^{T}x_S + C,
\end{equation}

where the effective bias vector becomes

\begin{equation}
b_S = 2Q_{SF}x_F .
\end{equation}

The reduced problem preserves global interaction effects while limiting dimensionality to a scale compatible with quantum execution.

\subsection{Quantum Optimization of Subproblems}

The reduced QUBO is solved using digitized counterdiabatic quantum optimization. Instead of exploring configurations through sequential local updates, quantum evolution samples correlated configurations within the active variable subspace. Because the bias term incorporates the influence of fixed variables, solutions obtained from the quantum optimizer remain consistent with the global objective.

Measurement outcomes generate candidate zoning updates, which are evaluated classically before incorporation into the global configuration.

\subsection{Iterative Hybrid Optimization Process}

Optimization proceeds through repeated interaction between classical coordination and quantum refinement. Beginning from an initial zoning configuration, impacts are computed for all variables and a new active subset is selected. Quantum optimization refines this subset, after which the global configuration is updated and impacts are recomputed.

As iterations progress, the location of active variables shifts across the network. Early iterations typically resolve large attribute imbalances, while later iterations refine spatial coherence along regional boundaries. This adaptive behavior emerges naturally from the impact definition rather than from externally imposed optimization stages.

The procedure terminates when successive iterations no longer produce meaningful reductions in objective value, indicating that dominant interaction conflicts have been resolved.

\subsection{Computational Perspective}

The proposed framework combines complementary strengths of classical and quantum computation. Classical processing evaluates global structure and maintains feasibility, whereas quantum optimization performs coordinated exploration within highly coupled subspaces. Because only a limited number of variables are optimized quantum mechanically at each step, the method scales to problem sizes substantially larger than available qubit counts while preserving the interaction structure of the original transportation model.

\section{Experiment Methodology and Validation Design}

The impact driven quantum decomposition framework introduced in the previous sections is derived from the interaction structure of the traffic zone QUBO formulation and from practical constraints imposed by current quantum hardware. The experimental study is therefore designed not only to evaluate solution quality, but also to validate whether the proposed algorithmic principles remain consistent under realistic execution conditions. In particular, the experiments examine three aspects of the framework: the ability of energy sensitivity to identify influential decision variables, the consistency between reduced quantum subproblems and the global objective, and the stability of hybrid optimization when executed on noisy intermediate-scale quantum devices.

\subsection{Synthetic Traffic Zone Environment}

A controlled traffic zoning environment is constructed to enable systematic observation of optimization behavior. The system consists of $n=64$ traffic zones arranged on a spatial grid with adjacency relationships defined between neighboring zones. This configuration preserves spatial coherence requirements typical of transportation planning while allowing clear visualization of partition outcomes.

Each zone is associated with multiple operational attributes representing heterogeneous transportation demand conditions. Attribute values are generated from bounded continuous distributions and normalized such that balanced partitioning corresponds to equal aggregate totals across operational regions. When translated into QUBO form, these attributes produce dense quadratic coupling among decision variables, reflecting long-range dependencies commonly observed in transportation systems.

A synthetic environment is intentionally adopted at this stage to isolate algorithmic behavior and ensure interpretability of optimization dynamics before deployment on real transportation datasets in future work.

\subsection{QUBO Construction}

The optimization objective follows directly from the formulation developed in Section III. A balance term penalizes deviation from equal regional workloads, while an adjacency term promotes spatial coherence among neighboring zones. The combined objective produces a dense QUBO matrix whose coefficients encode both global balancing interactions and local spatial constraints.

Because balance constraints introduce coupling among many zones simultaneously, the resulting problem exceeds the number of variables that can be reliably optimized in a single quantum execution. This property motivates the decomposition strategy proposed in Section IV.

\subsection{Quantum Hardware Environment}

\begin{figure}[t]
\centering
\includegraphics[width=0.48\textwidth]{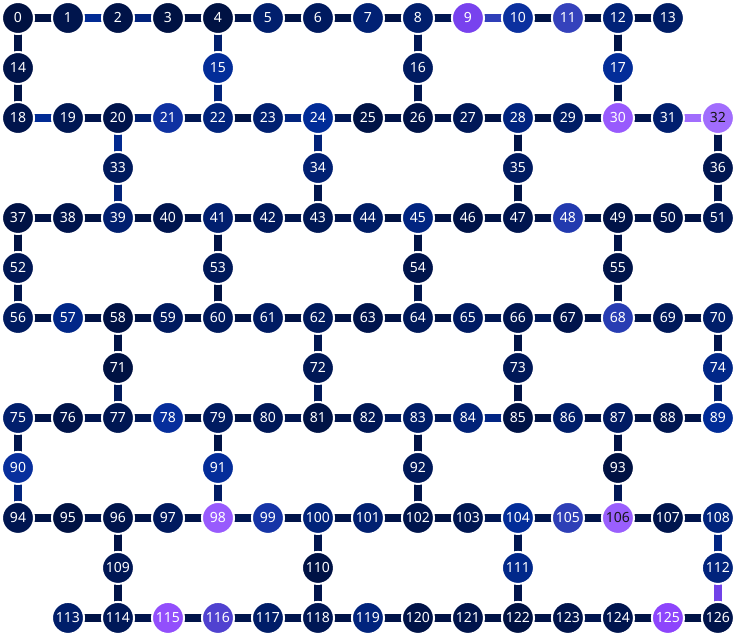}
\caption{Qubit connectivity layout and calibration characteristics of the IBM Quantum System One backend. Median readout and entangling gate error rates illustrate hardware noise levels that motivate reduced subQUBO optimization.}
\label{fig:ibm_layout}
\end{figure}

Quantum experiments were executed on the IBM Quantum System One device available through the \textit{IBM Quantum System One} backend. Figure~\ref{fig:ibm_layout} illustrates the qubit connectivity layout of the processor together with representative calibration metrics, including readout assignment error and two-qubit gate (ECR) error rates.

The topology shows limited qubit connectivity and nonuniform error distributions across the device (see Figure~\ref{fig:ibm_layout}). Median readout assignment errors on the order of $10^{-2}$ and entangling gate errors on the order of $10^{-3}$ introduce stochastic noise during optimization. These hardware characteristics restrict reliable circuit depth and effective problem size, providing a practical justification for solving reduced subQUBO instances instead of the full optimization problem.

\subsection{Validation Configurations}

To isolate the contribution of each design component, three optimization configurations are evaluated.

The first configuration applies direct quantum optimization to the full QUBO model, establishing a feasibility reference and demonstrating that transportation partitioning objectives can be executed within a gate-based quantum workflow.

The second configuration employs classical SubQUBO iterative refinement without impact guidance. This experiment evaluates whether decomposition alone improves optimization performance.

The third configuration implements the proposed impact driven hybrid framework. At each iteration, variable impacts derived from the global objective determine the active subset, a reduced QUBO is constructed, and quantum optimization refines the selected variables before reintegration into the global configuration.

Comparing these configurations enables validation of the hypothesis that energy sensitivity guided decomposition improves optimization stability under realistic hardware constraints.

\subsection{Evaluation Metrics}

Evaluation emphasizes transportation-relevant outcomes rather than computational speed. The primary quantitative metric is the QUBO objective value, which jointly measures attribute balance and spatial coherence. Tracking objective values across iterations enables analysis of convergence behavior and comparison among optimization strategies.

Spatial partitions are additionally examined to assess regional continuity and fragmentation, allowing qualitative interpretation of how optimization updates translate into transportation planning structures.

\subsection{Observation and Convergence Analysis}

Objective values are recorded after each iteration of the hybrid workflow to reconstruct optimization trajectories. This longitudinal observation allows investigation of whether impact based selection consistently targets high-conflict regions and whether quantum refinement produces monotonic improvement.

Because intermediate partitions are not explicitly optimized globally, convergence behavior provides an important validation signal for the hybrid design.

\subsection{Implementation Environment}

All classical components, including QUBO construction, impact computation, and iterative coordination, were implemented in Python. Quantum optimization was performed using the Iskay optimizer through the Qiskit runtime environment. Returned candidate solutions were incorporated directly into the hybrid optimization loop without additional heuristic tuning, ensuring that observed performance reflects the proposed algorithmic framework rather than postprocessing adjustments.
\section{Results and Analysis}

This section evaluates experimental outcomes using the validation framework defined in Section V. The analysis examines three aspects of algorithm behavior: feasibility of direct quantum optimization, limitations of classical iterative decomposition, and the effect of impact driven hybrid optimization. Results are interpreted based on observable optimization trajectories and resulting traffic zone partitions rather than claims of computational advantage.

\subsection{Direct Quantum Optimization}

The complete QUBO formulation was first solved using the Iskay optimizer without decomposition. The best observed solution achieved an objective value of

\begin{equation}
H = 151.1 .
\end{equation}

\begin{figure}[t]
\centering
\includegraphics[width=0.48\textwidth]{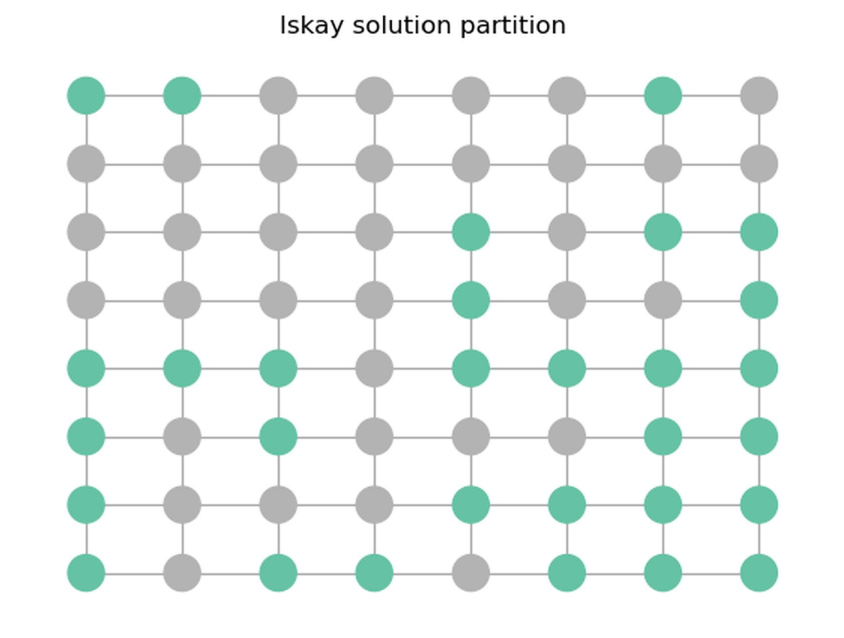}
\caption{Traffic zone partition obtained from direct Iskay optimization with objective value $H=151.1$.}
\label{fig:direct_partition}
\end{figure}

Figure~\ref{fig:direct_partition} presents the resulting traffic zone partition. The solution forms spatially coherent regions while maintaining balanced attribute distribution, confirming that the transportation zoning formulation can be executed directly within a gate-based quantum optimization workflow at this problem scale.

This experiment establishes a reference solution representing the lowest energy configuration observed in the study.

\subsection{Classical Iterative SubQUBO Baseline}

\begin{figure}[t]
\centering
\includegraphics[width=0.48\textwidth]{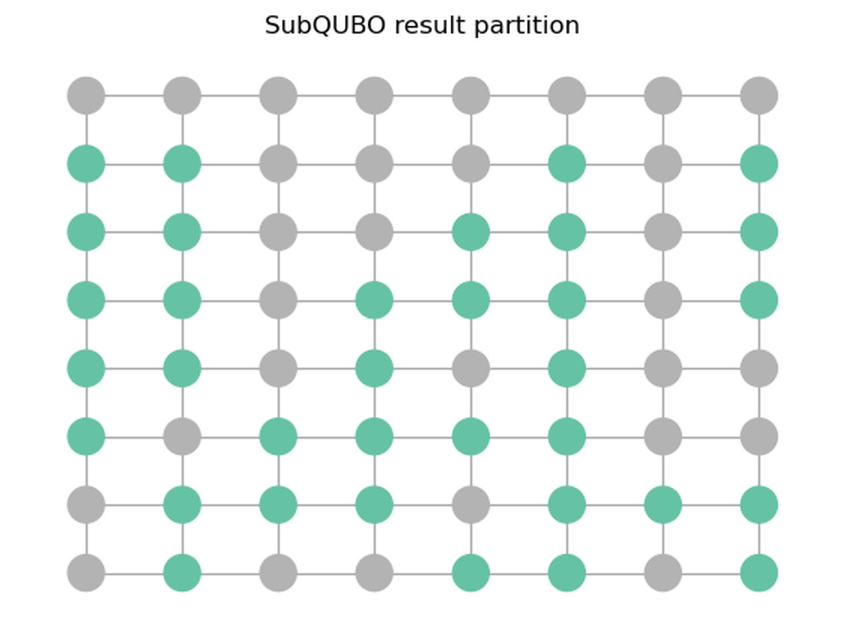}
\caption{Traffic zone partition produced by classical SubQUBO iterative refinement.}
\label{fig:classical_partition}
\end{figure}

\begin{figure}[t]
\centering
\includegraphics[width=0.48\textwidth]{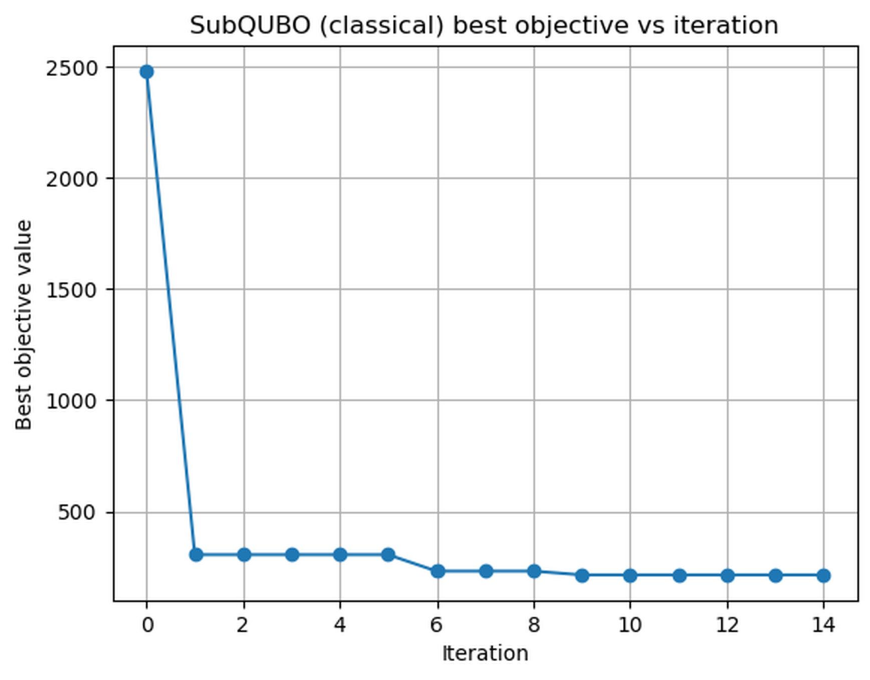}
\caption{Objective value trajectory for classical SubQUBO refinement, converging to $H=214.0$.}
\label{fig:classical_objective}
\end{figure}

A classical SubQUBO refinement strategy was then evaluated to isolate the effect of decomposition without quantum assistance. The corresponding partition is shown in Figure~\ref{fig:classical_partition}, and the objective trajectory is shown in Figure~\ref{fig:classical_objective}.

The optimization rapidly reduces objective value during the first iteration but soon stabilizes at

\begin{equation}
H = 214.0 ,
\end{equation}

indicating convergence to a local minimum. The resulting partition exhibits fragmented spatial structure compared with the direct quantum solution, suggesting that sequential subset updates fail to coordinate strongly coupled zones across the network.

This behavior confirms that decomposition alone does not resolve dense global interactions present in the transportation QUBO model.

\subsection{Impact Driven Hybrid Optimization}

The proposed impact driven decomposition framework was evaluated using the IBM Quantum System One backend. At each iteration, variables with the highest impact values were selected to form reduced QUBO instances optimized through quantum execution.

\begin{figure}[t]
\centering
\includegraphics[width=0.48\textwidth]{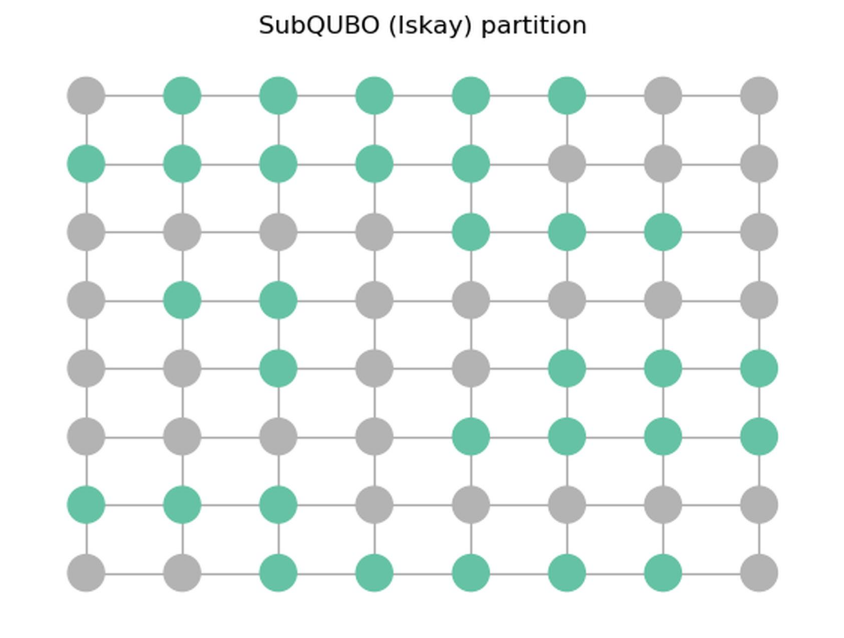}
\caption{Traffic zone partition obtained using impact driven hybrid optimization.}
\label{fig:hybrid_partition}
\end{figure}

\begin{figure}[t]
\centering
\includegraphics[width=0.48\textwidth]{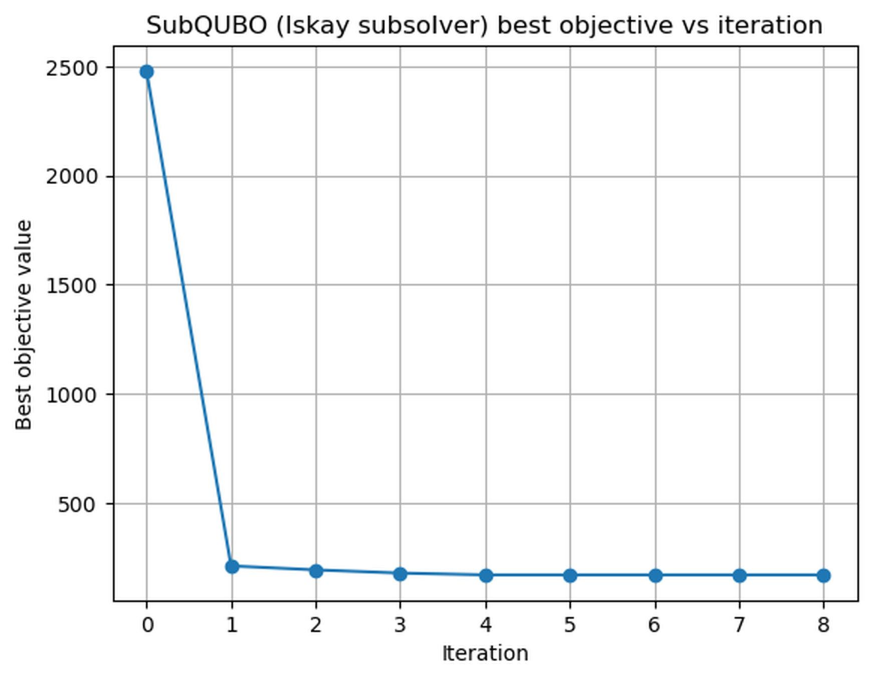}
\caption{Objective value trajectory during hybrid impact driven optimization.}
\label{fig:hybrid_objective}
\end{figure}

The final partition obtained from the hybrid procedure is shown in Figure~\ref{fig:hybrid_partition}, while the objective trajectory appears in Figure~\ref{fig:hybrid_objective}.

Starting from the classical baseline value of 214.0, the hybrid approach progressively reduces the objective to

\begin{equation}
H = 169.5 ,
\end{equation}

after four iterations. The convergence curve shows steady improvement followed by stabilization, indicating that impact based selection consistently identifies influential variables for refinement.

Compared with the classical partition, the hybrid solution exhibits improved spatial consistency and reduced fragmentation. However, the objective value remains above the direct quantum result of 151.1.

\subsection{Comparative Performance Summary}

Table~\ref{tab:performance_summary} summarizes outcomes across all configurations. The hybrid framework reduces the objective by approximately 44 points relative to the classical baseline, demonstrating that impact guided selection improves optimization effectiveness even when quantum execution is restricted to subproblems.

\begin{table}[t]
\caption{Comparison of optimization outcomes across configurations.}
\centering
\small
\setlength{\tabcolsep}{3pt}
\begin{tabular}{lcc}
\toprule
Method & Final objective & Outcome \\
\midrule
Direct quantum solve & 151.1 & lowest energy \\
Hybrid impact driven & 169.5 & improved search \\
Classical iterative & 214.0 & early stagnation \\
\bottomrule
\end{tabular}
\label{tab:performance_summary}
\end{table}

\subsection{Interpretation of Findings}

Several observations follow directly from the experimental results.

First, direct quantum optimization confirms feasibility of solving moderate-scale transportation QUBO models without decomposition.

Second, classical SubQUBO refinement alone exhibits early stagnation, indicating that unguided decomposition cannot adequately address dense interaction structures.

Third, impact driven hybrid optimization improves convergence behavior by directing quantum computation toward variables with the greatest influence on global energy. This produces consistent objective reduction and more coherent spatial partitions.

Finally, the hybrid method does not outperform the direct quantum solution. This suggests a tradeoff introduced by decomposition: while subproblem optimization enables scalability under hardware constraints, restricting optimization to subsets of variables may limit exploration of globally correlated configurations available in full problem execution.

The experiments validate the design logic of impact guided hybrid optimization while clarifying its practical limitations under current quantum hardware conditions. The direct quantum solution serves as a feasibility reference rather than a scalable baseline. As problem size increases, full QUBO execution becomes infeasible due to qubit count, connectivity, and noise constraints, motivating the hybrid strategy investigated in this work.

\section{Conclusion and Future Work}

This paper presented an impact-driven hybrid quantum–classical framework for traffic zone partitioning formulated as a QUBO optimization problem. By analyzing the interaction structure of transportation zoning objectives, the proposed method allocates quantum computation selectively to influential subsets of decision variables while maintaining global feasibility through classical coordination.

Experimental evaluation demonstrates several key findings. Direct quantum optimization successfully solves moderate-scale transportation QUBO instances, establishing feasibility of gate-model quantum execution for zoning problems. Classical SubQUBO refinement alone exhibits early stagnation, indicating that decomposition without guidance fails to resolve dense global interactions. The proposed impact-driven hybrid approach improves convergence behavior and produces more spatially coherent partitions relative to classical refinement, confirming the effectiveness of energy sensitivity as a decomposition principle.

At the same time, the hybrid method does not outperform the best direct quantum solution. This observation highlights an inherent tradeoff introduced by decomposition: restricting optimization to subspaces improves scalability under hardware constraints but may limit exploration of globally correlated configurations. The results therefore suggest that hybrid quantum optimization should be viewed as a scalability mechanism rather than a direct performance replacement for full problem execution.

Future work will investigate several directions. First, adaptive selection of subproblem size based on hardware calibration metrics may improve robustness under varying noise conditions. Second, integration with larger transportation datasets will allow evaluation of real-world zoning applications. Third, advances in qubit connectivity and error mitigation techniques may enable tighter coupling between subQUBO solutions and global optimization dynamics. Finally, extending the framework toward multi-region partitioning and dynamic transportation environments represents an important step toward practical deployment of hybrid quantum optimization in intelligent transportation systems.

\section{Acknowledgment}
This research is funded through the IBM-RPI Future of Computing Collaboration.
	\bibliographystyle{IEEEtran}
	\bibliography{root} 
    
\end{document}